\documentclass{SciPost}

\hypersetup{
    colorlinks,
    linkcolor={red!50!black},
    citecolor={blue!50!black},
    urlcolor={blue!80!black}
}

\usepackage[bitstream-charter]{mathdesign}
\DeclareSymbolFont{usualmathcal}{OMS}{cmsy}{m}{n}
\DeclareSymbolFontAlphabet{\mathcal}{usualmathcal}

\fancypagestyle{SPstyle}{
\fancyhf{}
\lhead{\raisebox{-1.5mm}[0pt][0pt]{\href{https://scipost.org}{\includegraphics[width=20mm]{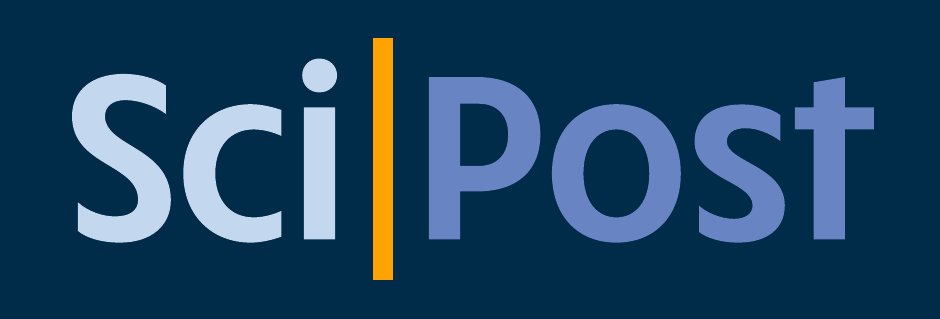}}}}

\rhead{\small \href{https://scipost.org/SciPostPhysProc.18.014}{SciPost Phys. Proc. 18, 014 (2026)}}

\fancyfoot[C]{\textbf{014.\thepage}}
}

\newcommand\abstr[1]{{\boldmath\textbf{#1}}}

\usepackage{xspace}
\usepackage{subfig}
\usepackage{verbatim}
\usepackage{booktabs}
\usepackage{array}
\usepackage{tabularx}

\newcolumntype{C}[1]{>{\centering\arraybackslash}p{#1}}
\newcommand{\loqcd}{\rm LO_{1}\xspace}

\newcommand{\nloone}{{\rm NLO}_1\xspace}
\newcommand{\nlotwo}{{\rm NLO}_2\xspace}
\newcommand{\nlothree}{{\rm NLO}_3\xspace}

\def\refeq#1{\mbox{Eq.~(\ref{#1})}}

\def\reffi#1{\mbox{Fig.~\ref{#1}}}

\def\refta#1{\mbox{Table~\ref{#1}}}

\def\citere#1{\mbox{Ref.~\cite{#1}}}

\newcommand{\newc}{\newcommand}
\newc{\beq}{\begin{equation}}
\newc{\eeq}{\end{equation}}

\newc{\bit}{\begin{itemize}}
\newc{\eit}{\end{itemize}}
\newc{\ben}{\begin{enumerate}}
\newc{\een}{\end{enumerate}}
\newc{\bce}{\begin{center}}
\newc{\ece}{\end{center}}
\newc{\bfi}{\begin{figure}}
\newc{\efi}{\end{figure}}

\newcommand{\rT}{{\mathrm{T}}}

\newcommand{\TeV}{\ensuremath{\,\text{TeV}}\xspace}

\newcommand{\Pp}{\ensuremath{\text{p}}}
\newcommand{\Pe}{\ensuremath{\text{e}}\xspace}
\newcommand{\Pb}{\ensuremath{\text{b}}\xspace}
\newcommand{\Pt}{\ensuremath{\text{t}}\xspace}

\newcommand{\PZ}{\ensuremath{\text{Z}}\xspace}

\newcommand{\recola}{{\sc Recola}\xspace}

\newcommand{\mocanlo}{{\sc MoCaNLO}\xspace}
\newcommand{\collier}{{\sc Collier}\xspace}

\newcolumntype{.}{D{.}{.}{-1}}
\newcolumntype{d}[1]{D{.}{.}{#1}}
\colorlet{tableoverheadcolor}{gray!37.5}
\colorlet{tableheadcolor}{gray!25}
\colorlet{tablerowcolor}{gray!12.5}

\newcommand{\tm}[1]{M_{\rT,{#1}}}

\makeatletter
\g@addto@macro\bfseries{\boldmath}
\makeatother

\begin{document}

\pagestyle{SPstyle}

\begin{center}{\Large \textbf{\color{scipostdeepblue}{
Predictions for off-shell $t\bar{t}Z$ production at the LHC:\\The complete set of LO and NLO contributions\\
}}}\end{center}

\begin{center}
\textbf{
Daniele Lombardi
}
\end{center}

\begin{center}
Universit\"at W\"urzburg, Institut f\"ur Theoretische Physik\\und Astrophysik,
97074 W\"urzburg, Germany
\\[\baselineskip]
\href{mailto:daniele.lombardi@uni-wuerzburg.de}{\small daniele.lombardi@uni-wuerzburg.de}

\end{center}

\definecolor{palegray}{gray}{0.95}
\begin{center}
\colorbox{palegray}{
  \begin{tabular}{rr}
  \begin{minipage}{0.4\textwidth}
    \includegraphics[width=\textwidth]{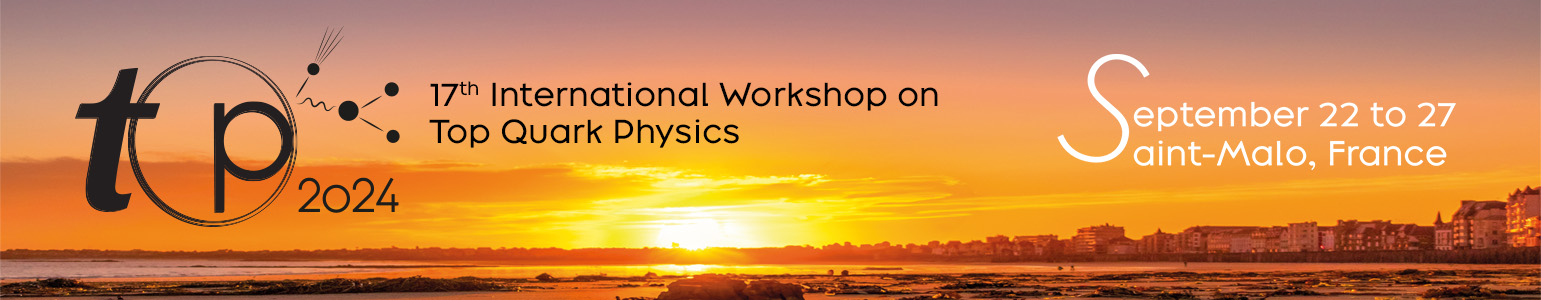}
  \end{minipage}
  &
  \begin{minipage}{0.48\textwidth}
    \begin{center}
        {\it The 17th International Workshop on\\ Top Quark Physics (TOP2024)} \\
        {\it Saint-Malo, France, 22-27 September 2024}
        \doi{10.21468/SciPostPhysProc.18}\\
    \end{center}
  \end{minipage}
\end{tabular}
}
\end{center}

\section*{\color{scipostdeepblue}{Abstract}}
\abstr{%
We present predictions for the production and decay of a top--antitop pair in association
with a $\rm Z$ boson  in the multi-lepton decay channel at the LHC. Our results include
the complete set of LO and NLO contributions. Since our calculation is based on full matrix
elements, off-shell effects are entirely taken into account.
Integrated and differential cross-sections are reported for a realistic fiducial setup.
}

\begin{center}
\begin{tabular}{lr}
\begin{minipage}{0.56\textwidth}
\raisebox{-1mm}[0pt][0pt]{\includegraphics[width=12mm]{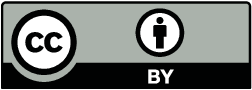}}
{\small Copyright D. Lombardi. \newline
This work is licensed under the Creative Commons \newline
\href{http://creativecommons.org/licenses/by/4.0/}{Attribution 4.0 International License}. \newline
Published by the SciPost Foundation.
}
\end{minipage}
&
\begin{minipage}{0.44\textwidth}
    \noindent\begin{minipage}{0.68\textwidth}
    {\small Received 2024-10-21 \newline Accepted 2024-12-18 \newline Published 2026-01-29}
    \end{minipage}
    \begin{minipage}{0.25\textwidth}
    \begin{center}
    \href{https://crossmark.crossref.org/dialog/?doi=10.21468/SciPostPhysProc.18.014&amp;domain=pdf&amp;date_stamp=2026-01-29}{\includegraphics[width=7mm]{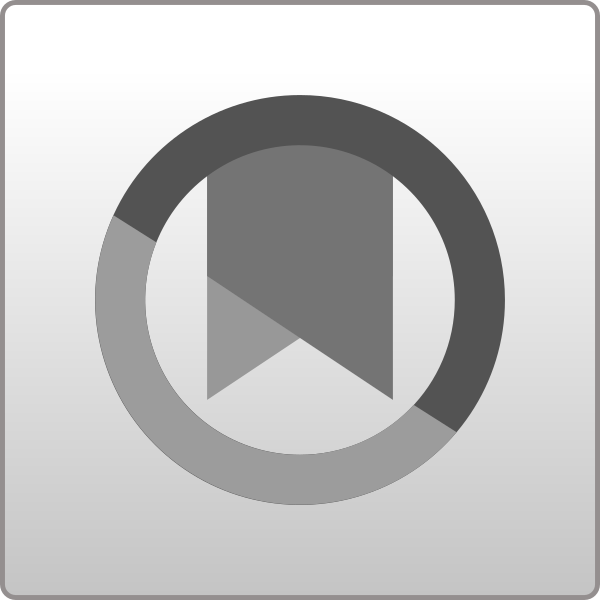}}\\
    \tiny{Check for}\\
    \tiny{updates}
    \end{center}
    \end{minipage}
    \\\\
    \small{\doi{10.21468/SciPostPhysProc.18.014}}
\end{minipage}
\end{tabular}
\end{center}

\vspace{10pt}
\noindent\rule{\textwidth}{1pt}

\section{Introduction}
\label{sec:intro}
The relevance of the top-quark-associated $\PZ$-boson production ($\rm t\bar{\rm t}Z$ in the following)
is nowadays well established and confirmed by the numerous searches that have been performed by
experimental collaborations in the course of the Large Hadron Collider (LHC) life time. Indeed, its indepth investigation
 can offer for instance additional tests on the Standard Model (SM), better control on the background
to other processes, and a direct access to the top-quark couplings with the electroweak (EW) sector. 

For all these reasons, the theory community has tried hard in the last few decades to continuously
improve the accuracy of the predictions for $\rm t\bar{\rm t}Z$ production.
This problem has been tackled for a long time with simplified approaches, where typically resonance
particles are set on shell or modelled with the narrow-width approximation.
Only quite recently an off-shell calculation for $\rm t\bar{\rm t}Z$ in the four-charged-lepton decay
channel came out in \citere{Bevilacqua:2022nrm},
where it was shown how crucial a correct
modelling of the off-shell effects can be, especially for some differential observables. In that
work, NLO accuracy in QCD was achieved. 
In this contribution, we discuss some results from \citere{Denner:2023eti}, where
all still-missing LO and NLO terms have been obtained with a full off-shell
calculation. The same results were already presented in \citere{Lombardi:2024owo}.

\section{Outline of the calculation}
\begin{figure}[t]
  \centering
  \includegraphics[scale=0.27]{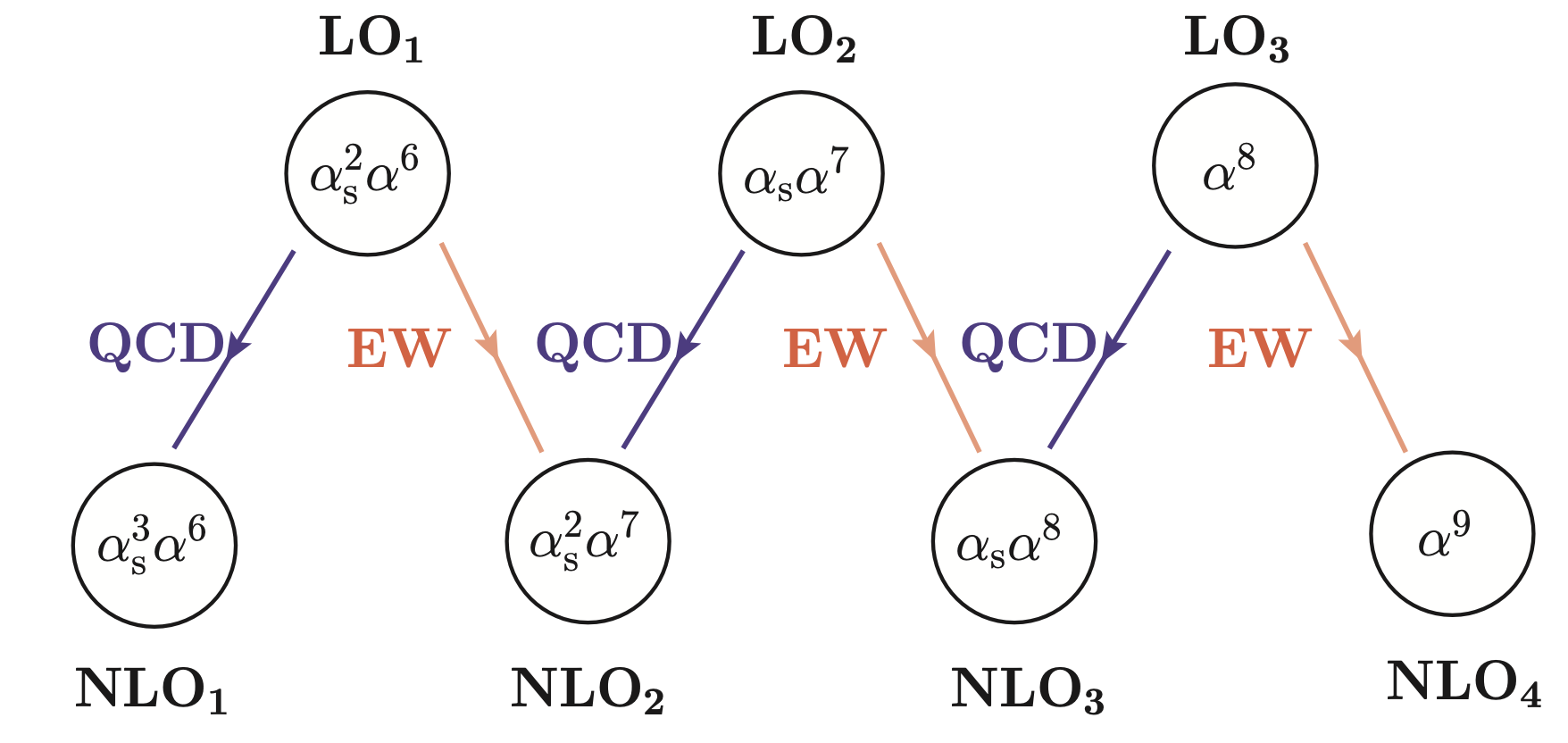}
  \caption{
    Perturbative orders contributing at LO and NLO
    for the process in \refeq{eq:procdef}.
  }\label{fig:orders}
\end{figure}

In \citere{Denner:2023eti} the different LO contributions and their NLO EW and QCD corrections to the process
\begin{equation}
  \label{eq:procdef}
  \Pp\Pp\to\Pe^+\nu_\Pe\,\mu^-\overline{\nu}_\mu\,{\Pb}\,\overline{\Pb}\,{\tau}^+{\tau}^-\,,
\end{equation}
have been computed using the private Monte Carlo generator \mocanlo. The latter program 
proved to be particularly efficient in dealing with processes with an intricate resonance structure,
thanks to its highly-optimised  multichannel integration. The code is interfaced with
\recola \cite{Actis:2012qn, Actis:2016mpe}  and \collier \cite{Denner:2016kdg}, which allow
for the numerical evaluation of all necessary tree-level SM matrix elements and  one-loop integrals.
A realistic modelling of
off-shell effects is achieved by including the complete set of resonant and non-resonant Feynman
diagrams. All relevant partonic channels have been considered, together with the photon- and
bottom-induced ones. That has been done systematically for each of the perturbative orders
contributing to the process (which are schematically summarized in \reffi{fig:orders}).

\section{Results}
\label{sec:results}

Our results have been obtained at a centre-of-mass energy of 13\TeV in the fiducial region described
in \citere{Denner:2023eti}. Most notably, the infrared safety of the calculation when three bottom
quarks occur in the final state has been ensured by proper recombination rules as part of the
jet-clustering algorithm, where a $\Pb$~jet and a light jet are recombined into a $\Pb$~jet, and
two $\Pb$~jets into a light jet. The factorization and renormalizaton scales have been set to
$\mu_0 =\frac{1}{2}{\left(\tm{\Pt}\,\tm{\overline{\Pt}}\right)}^{1/2}$,
which is based on the transverse masses of the reconstructed
top/anti-top quarks. Theory uncertainties have been estimated with the standard
$7$-point scale variation.

\subsection{Integrated cross-sections}
\label{sec:integrated_xs}
\begin{table}[t]
\renewcommand{\arraystretch}{1.4}
\centering
\caption{ 
LO cross-sections and NLO corrections (in ab) in the fiducial setup. In the second column all partonic channels are included in $\sigma_{\rm{no}\Pb}$
except the ones having at least one bottom quark in the initial
state, while  $\sigma_{\Pb}$ includes all these channels.
The sum of the two ($\sigma$) is shown in the sixth column. Ratios with respect to the cross-section $\sigma_{\rm{no}\Pb}$ at
$\loqcd$ accuracy are reported in the third and fifth column. In the seventh column ratios are shown with respect to the full
$\loqcd$ cross-section including the bottom channels, as
well. Integration errors are given in parentheses and percentage 7-point scale
variations as super- and sub-scripts. See \citere{Lombardi:2024owo}.}\label{table:sigmainclNLO_combined}
\resizebox{\textwidth}{!}{
\begin{tabular}{c|cr|cr|cr}
  \hline
\multicolumn{1}{c|}{\textrm{perturbative order}} & $\sigma_{\rm{no}\Pb}$
[ab]   &  
$\frac{\sigma_{\rm{no}\Pb}}{\sigma_{\rm{no}\Pb,\,\rm{LO}_{1}}}$ & $\sigma_{\Pb}$ [ab]   &  $\frac{\sigma_{\Pb}}{\sigma_{\rm{no}\Pb,\,\rm{LO}_{1}}}$& $\sigma$ [ab]   &  $\frac{\sigma}{\sigma_{\rm{LO}_{1}}}$  \\ 
\hline
$\rm LO_{1}$  &  107.246(5)$^{+35.0\%}_{-24.0\%}$  &  1.0000 & 0.31378(9)& $+$0.0029 &107.560(5)$^{+34.9\%}_{-23.9\%}$& 1.0000\\[0.9ex]
$\rm LO_{2}$  &  0.7522(2)$^{+11.1\%}_{-9.0\%}$ & $+$0.0070  & $-$0.6305(2)& $-$0.0059 &0.1217(3)& $+$0.0011\\[0.9ex]
$\rm LO_{3}$  &  0.2862(1)$^{+3.4\%}_{-3.4\%}$  &
$+$0.0027&0.7879(2)&$+$0.0073 & 1.0742(3)$^{+12.1\%}_{-14.9\%}$& $+$0.0100 \\[0.9ex]
  \hline
${\rm NLO_1}$  & $-$11.4(1)    & $-$0.1072  & 0.518(3)    & $+$0.0048 &$-$10.9(1)& $-$0.1016\\ 
  ${\rm NLO_2}$  & $-$0.89(1)    & $-$0.0083  & 0.109(3)  & $+$0.0010 & $-$0.78(1) & $-$0.0072\\
${\rm NLO_3}$  & 1.126(4)      & $+$0.0105  & $-$0.089(4) & $-$0.0008 & 1.037(6)   & $+$0.0096\\
${\rm NLO_4}$  &  $-$0.0340(9) & $-$0.0003  & $-$0.0180(9)& $-$0.0002 & $-$0.052(1) & $-$0.0005\\
  \hline
$\rm LO_{1}$+${\rm NLO_1}$ & 95.8(1)$^{+0.4\%}_{-11.2\%}$ &$+$0.8933&0.832(3)& $+$0.0078&96.6(1)$^{+0.4\%}_{-10.7\%}$& $+$0.8984\\[0.9ex]
  \hline
  \hline
    LO & 108.285(5)$^{+34.7\%}_{-23.8\%}$  &$+$1.0097  &0.4713(3) & $+$0.0044 & 108.756(5)$^{+34.5\%}_{-23.7\%}$& $+$1.0111\\[0.9ex]
    LO+NLO & 97.0(1)$^{+0.5\%}_{-11.2\%}$  & $+$0.9052  &0.991(6) &$+$0.0092 & 98.0(1)$^{+0.4\%}_{-10.6\%}$& $+$0.9114\\[0.9ex]
  \hline
\end{tabular}}
\end{table} 

\refta{table:sigmainclNLO_combined} reports results for the integrated cross-sections at different
perturbative accuracies, both without and including the bottom-channel contribution in
$\sigma_{\rm{no}\Pb}$ and $\sigma$, respectively. One can see that QCD corrections of $\nloone$ type
are the dominant ones, with a $-10\%$ K-factor and causing a large reduction in the size of the $\loqcd$ uncertainty
bands. The additional LO and NLO subleading contributions, which mostly cancel against each other,
just correct the $\loqcd$ result at the sub-percent level. We also observe that the inclusion
of bottom-induced channels has a moderate impact of $+1\%$ on the full NLO prediction. 

\subsection{Differential cross-sections}
\label{sec:differential_xs}
\begin{figure}[t]
  \centering
  \subfloat[Transverse momentum of the bottom-jet pair.\label{fig:ptbb}]{  \includegraphics[scale=0.27]{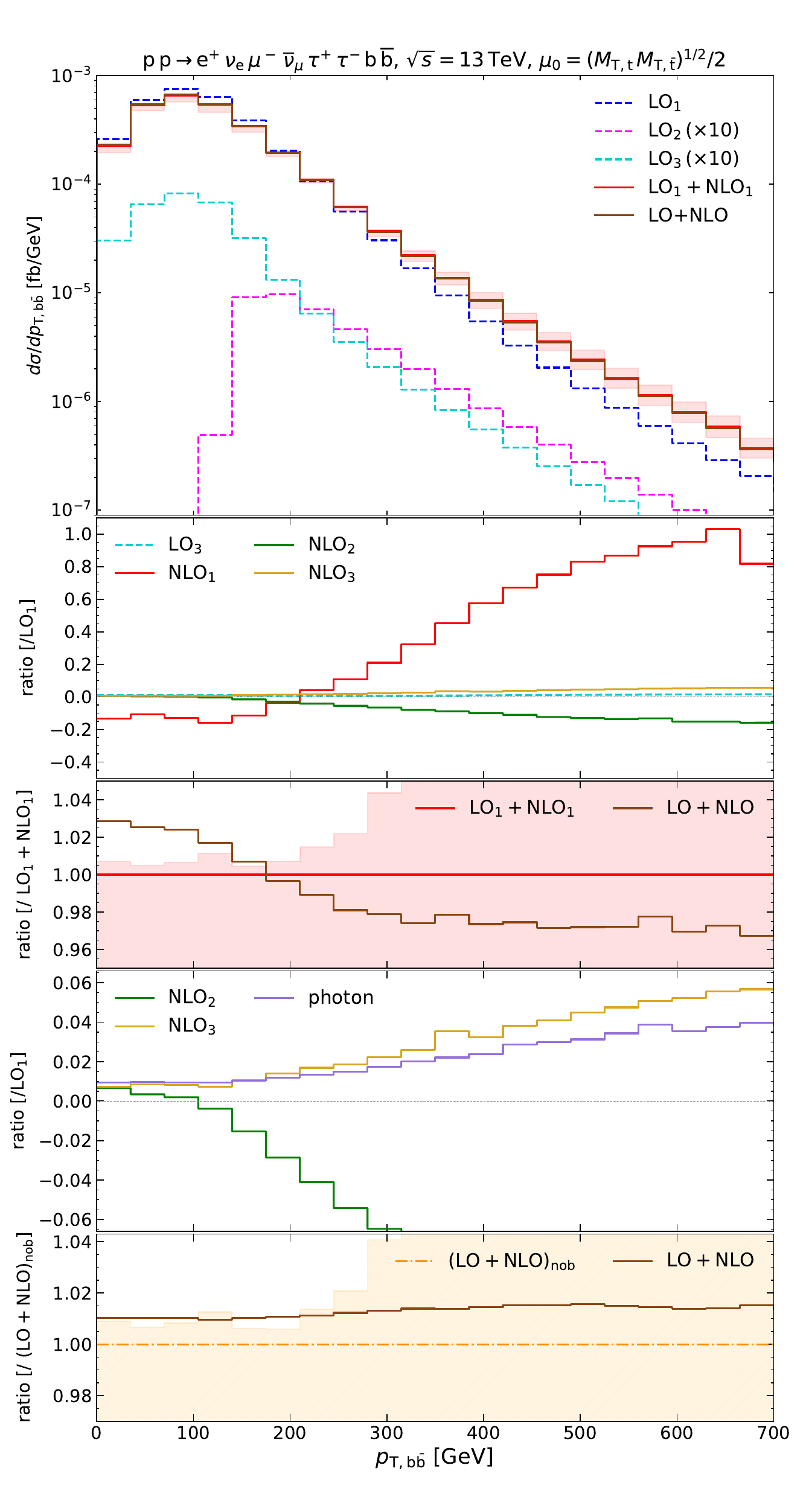}}
  \subfloat[Invariant mass of the $\tau^+\tau^-$ pair.\label{fig:mtt}]{  \includegraphics[scale=0.27]{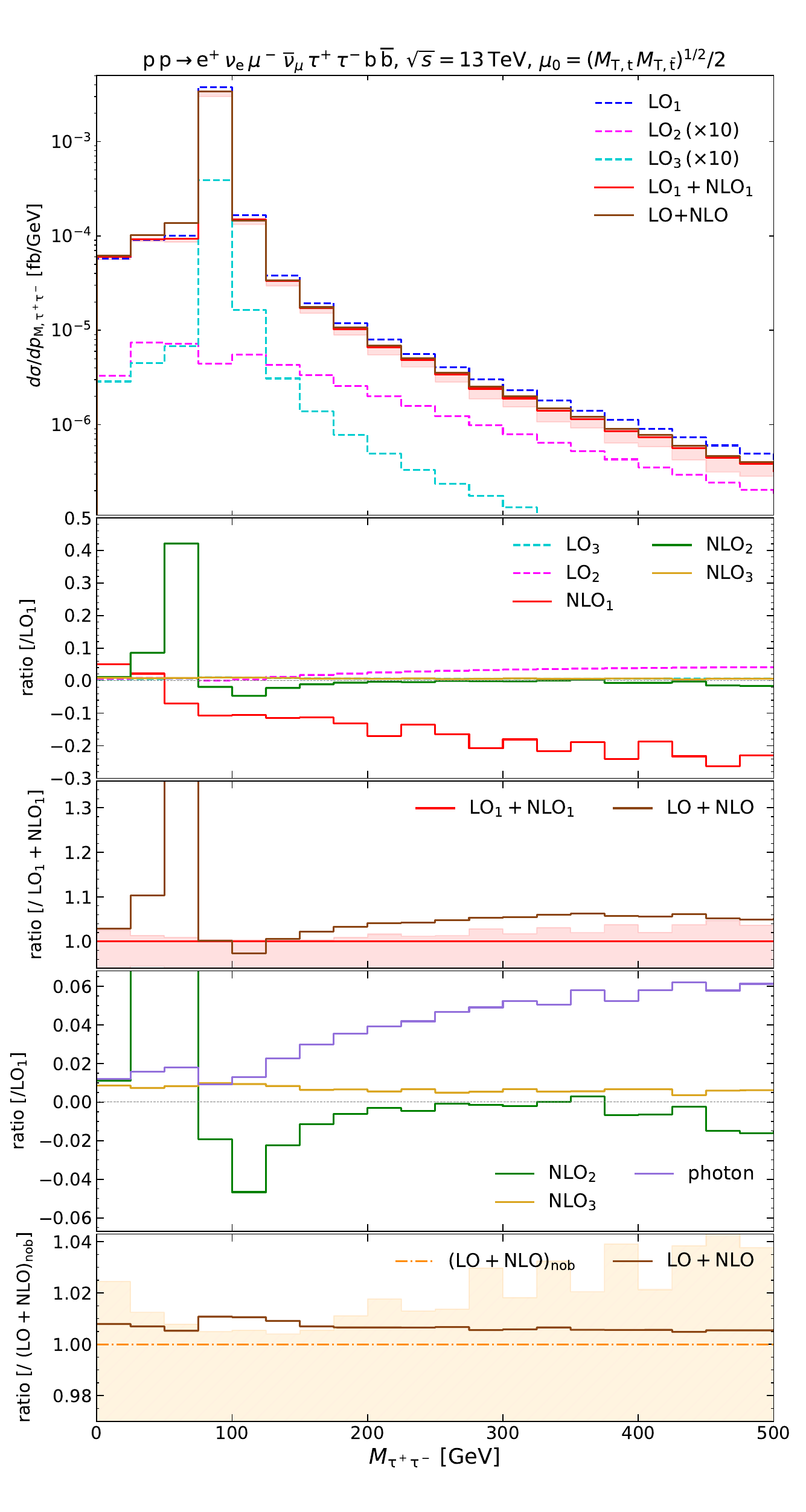}}
    \caption{
Distributions in the transverse momentum of the bottom-jet pair (left) and in the invariant mass of the $\tau^+\tau^-$ pair (right).
    The different NLO corrections for the observables are compared  separately (first ratio panels) and at the level of the full prediction (second ratio panel).
    The size of photon-induced channels and bottom contributions are presented in the third and fourth ratio panels, respectively. See \citere{Lombardi:2024owo}.
  }\label{fig:diff4}
\end{figure}

\reffi{fig:diff4} shows  the interplay of the different perturbative
contributions for two illustrative observables.
The distribution in transverse momentum of the $\Pb\bar\Pb$ pair in \reffi{fig:ptbb}
is extremely sensitivity in its tails to QCD corrections (in red), that manifest as a giant QCD K-factor.
Nevertheless, subleading contributions provide important corrections to the observable in the same
phase-space region: the $\nlotwo$ term (in green) can reach up to $-15\%$ of the $\loqcd$ at
$p_{\rm T,\Pb\bar\Pb}=700\,$GeV, due to large EW Sudakov logarithms, and the $\nlothree$ term
(in goldenrod) a positive $+5\%$. Even if the definition of this observable requires two $\Pb$ jets,
the inclusion of bottom-induced channels just corresponds to a small normalisation effect over the full
spectrum.
Also for the invariant mass of the $\tau^+\tau^-$ pair in \reffi{fig:mtt}
accounting for perturbative orders beyond $\loqcd$ and $\nloone$ becomes crucial. The $\nloone$ dominates in
the far off-shell region with a $-20\%$ correction to the $\loqcd$. In the same region, due to the
growth of the photon PDF, photon contributions (in violet) turn out to be the most important subleading terms,
with a K-factor up to $+6\%$. Finally, the photon radiative effects
included in the $\nlotwo$ term are the most significant ones right below the $\PZ$-boson
resonance peak, where they  amount to a $+40\%$ correction.

\section{Conclusion}
We have presented results for $\rm t\bar{\rm t}Z$ production for $13\,$TeV proton--proton collisions
that include the complete set of LO and NLO contributions and fully account for off-shell effects.
Even if the impact of subleading corrections is moderate at the integrated level, they can cause significant
shape distorsions for many observables. That renders their inclusion at the differential
level crucial in view of precise phenomenology and a realistic description of the process.

\section*{Acknowledgments} 
\paragraph{Funding information}
 This contribution is based upon work supported by the German Federal Ministry for
 Education and Research (BMBF) under contract no.~05H21WWCAA.

\end{document}